\PassOptionsToPackage{usenames,svgnames}{xcolor}

\documentclass[a4, 11pt, parskip=half, DIV=10]{scrartcl}
\usepackage[english]{babel}
\usepackage[T1]{fontenc}
\usepackage[utf8]{inputenc}

\usepackage{graphicx}
\usepackage[numbers]{natbib}
\usepackage{subcaption}
\usepackage{booktabs}
\usepackage{url}
\usepackage{todonotes}
\usepackage{multirow}
\usepackage{csquotes}
\usepackage{paralist}

\usepackage{pgfplots}
\usepgfplotslibrary{colorbrewer}

\pgfdeclarelayer{background}
\pgfsetlayers{background,main}

\usepackage{bm}

\usepackage{amsmath}
\usepackage{amssymb}
\usepackage{amsthm}
\usepackage{amsfonts}
\usepackage{thmtools}		
\usepackage{mleftright}
\usepackage{stmaryrd}
\usepackage{nicefrac}

\usepackage{algorithm}
\usepackage{algpseudocode}

\theoremstyle{definition}
\newtheorem{theorem}{Theorem}
\newtheorem{proposition}[theorem]{Proposition}

\newtheorem{lemma}{Lemma}

\newtheorem{definition}{Definition}

\usepackage{thm-restate}
\usepackage[mathic=true]{mathtools}
\usepackage{fixmath}
\usepackage{siunitx}
\usepackage{color}

\usepackage{pifont}

\usepackage{enumitem}
\setlist[enumerate]{itemsep=0.2ex, topsep=0.5\topsep}
\setlist[description]{itemsep=0.2ex, topsep=0.5\topsep}
\setlist[itemize]{itemsep=0.2ex, topsep=0.5\topsep}

\usepackage{setspace}
\usepackage{ellipsis}
\usepackage{xspace}
\usepackage{hfoldsty}

\usepackage{tcolorbox}
\usepackage{lmodern}
\usepackage[tt=false]{libertine}
\usepackage[varqu]{zi4}
\usepackage[libertine]{newtxmath}
\usepackage{abstract}
\usepackage{doi}

\makeatletter
\def\thmt@refnamewithcomma #1#2#3,#4,#5\@nil{%
	\@xa\def\csname\thmt@envname #1utorefname\endcsname{#3}%
	\ifcsname #2refname\endcsname
	\csname #2refname\expandafter\endcsname\expandafter{\thmt@envname}{#3}{#4}%
	\fi
}
\makeatother
\usepackage[capitalise,noabbrev]{cleveref}   

\DeclarePairedDelimiter{\norm}{\lVert}{\rVert}

\newcommand{\cNP}[0]{\ensuremath{\mathsf{NP}}}

\newcommand{\LLB}{\textit{LLB}}
\newcommand{\GED}{\textit{GED}}
\newcommand{\DLB}{\textit{DLB}}
\newcommand{\CLB}{\textit{CLB}}

\newcommand{\SLF}{\textit{SLF}}
\newcommand{\BranchLB}{\textit{BranchLB}}
\newcommand{\DB}{\ensuremath{\text{DB}}}
\newcommand{\kNN}[2]{\ensuremath{\operatorname{NN}(#1, #2)}}
\newcommand{\range}[2]{\ensuremath{\operatorname{range}(#1, #2)}}
\newcommand{\Gone}{G}
\newcommand{\Gtwo}{H}

\usepackage{lipsum}

\usepackage[auth-lg]{authblk}
\let\abs\relax

\recalctypearea

\begin{document}
\title{\vspace{-25pt} \text{EmbAssi}: Embedding Assignment Costs for Similarity Search in Large Graph Databases\thanks{This work was supported by the Vienna Science and Technology Fund (WWTF) through project VRG19-009. Additional funding was provided by the German Research Foundation (DFG) within the Collaborative Research Center SFB 876 \textit{Providing Information by Resource-Constrained Data Analysis}, DFG project number 124020371, SFB projects A2 and A6, \url{http://sfb876.tu-dortmund.de}.
\begin{center}
		\framebox{ \parbox{\textwidth-4\fboxsep}{This version of the article has been accepted for publication, after peer review but is not the Version of Record and does not reflect post-acceptance improvements, or any corrections. The Version of Record is available online at: https://doi.org/10.1007/s10618-022-00850-3}}
		\end{center}
}}

\author[1,2]{Franka Bause}
\author[3]{Erich Schubert}
\author[1,4]{Nils M.~Kriege}

\affil[1]{Faculty of Computer Science, University of Vienna, Vienna, Austria}
\affil[ ]{\ttfamily\{franka.bause,nils.kriege\}@univie.ac.at}
\affil[2]{UniVie Doctoral School Computer Science, University of Vienna, Vienna, Austria}
\affil[3]{Faculty of Computer Science, TU Dortmund University, Dortmund, Germany}
\affil[ ]{\ttfamily erich.schubert@tu-dortmund.de}
\affil[4]{Research Network Data Science, University of Vienna, Vienna, Austria}
\date{\vspace{-50pt}}

\maketitle

\begin{abstract}
\unboldmath%
The graph edit distance is an intuitive measure to quantify the dissimilarity of graphs, but its computation is \cNP-hard and challenging in practice.
We introduce methods for answering nearest neighbor and range queries regarding this distance efficiently for large databases with up to millions of graphs. We build on the filter-verification paradigm, where lower and upper bounds are used to reduce the number of exact computations of the graph edit distance. Highly effective bounds for this involve solving a linear assignment problem for each graph in the database, which is prohibitive in massive datasets.
Index-based approaches typically provide only weak bounds leading to high computational costs verification.
In this work, we derive novel lower bounds for efficient filtering from restricted assignment problems, where the cost function is a tree metric. 
This special case allows embedding the costs of optimal assignments isometrically into $\ell_1$ space, rendering efficient indexing possible. We propose several lower bounds of the graph edit distance obtained from tree metrics reflecting the edit costs, which are combined for effective filtering. Our method termed \textit{EmbAssi} can be integrated into existing filter-verification pipelines as a fast and effective pre-filtering step. Empirically we show that for many real-world graphs our lower bounds are already close to the exact graph edit distance, while our index construction and search scales to very large databases.
\end{abstract}
\subsection*{Keywords}
graph edit distance, similarity search, index

\section{Introduction}
\label{sec:introduction}

In various applications such as cheminformatics~\cite{Hernandez2019}, bioinformatics~\cite{Stoecker2019}, computer vision~\cite{xiao2013graph} and social network analysis, complex structured data arises, which can be naturally represented as graphs. To analyze large amounts of such data, meaningful measures of (dis)similarity are required. 
A widely accepted approach is the \emph{graph edit distance}, which measures the dissimilarity of two graphs in terms of the total cost of transforming one graph into the other by a sequence of edit operations. This concept is appealing because of its intuitive and comprehensible definition, its flexibility to adapt to different types of graphs and annotations, and the interpretability of the dissimilarity measure. However, computing the graph edit distance for a pair of graphs is \cNP-hard~\cite{2_ComparingStars} and challenging in practice even for small graphs. 
This renders similarity search regarding the graph edit distance in databases difficult, which is relevant in many applications.
A prime example is a molecular information system, which often contains millions of graphs representing small molecules. A standard task in computational drug discovery is similarity search in such databases, for which the concept of graph edit distance has proven useful~\cite{Hernandez2019}.
However, the extensive use of graph-based methods in such systems is still hindered by the computational burden, especially in comparison to embedding-based techniques, for which efficient similarity search is well studied~\cite{Ramzi2010}.
Moreover, similarity search is the fundamental problem when using the graph edit distance in downstream supervised or unsupervised machine learning methods such as $k$-nearest neighbors classification. Promising results have been reported for classifying graphs from diverse applications representing, e.g., small molecules~\cite{1_gedLinear}, petroglyphs~\cite{petroglyphs}, or cuneiform signs~\cite{cuneiform}. However, this approach does not readily scale to large datasets, where embedding-based methods such as graph kernels~\cite{gk_survey} and graph neural networks~\cite{gnn_survey} have become the dominating techniques.

Algorithms for exact~\cite{Gouda2016,34_blp,Chang2020,CHEN2019762} or approximate~\cite{35_beamS,27_Riesen,1_gedLinear} graph edit distance computation have been extensively studied.
They are typically optimized for pairwise comparison but can be accelerated in cases when a distance cutoff is given as part of the input.
While not directly suitable for searching large databases, these algorithms are used in the verification step after a set of candidates has been obtained by filtering.
In the filtering step, lower bounds on the graph edit distance are typically used to eliminate graphs that cannot satisfy the distance threshold, while upper bounds are used to add graphs to the answer set without the need for verification.
Several techniques following this paradigm have been proposed, see Table~\ref{table:relatedWork} for an overview.
An important characteristic for scalability is whether these techniques use an index to avoid scanning every graph in the database.
This is not directly possible for many of the existing bounds on the graph edit distance, which were often studied in other contexts.
A recent systematic comparison of existing bounds~\cite{29_ged_heuristics} shows that there is a trade-off between the efficiency of computation and the tightness of lower and upper bounds.
Lower bounds based on linear programming relaxations and solutions of the linear assignment problem were found to be most effective.
However, the computation of such bounds requires solving non-trivial optimization problems and is inefficient compared to computing standard distances on vectors. 
Moreover, their combination with well-studied indices for vector or metric data is often not feasible because they do not satisfy the necessary properties such as being metric or embeddable into vector space.
Qin et al.~\cite{QinBS20} concluded, that methods without an index do not scale well to very large databases, while those with an index often provide only loose bounds leading to a high computational cost for verification. To overcome this, they proposed an inexact filtering mechanism based on hashing, which cannot guarantee a complete answer set.
We show that exact similarity search in very large databases using the filter-verification paradigm is possible.
We achieve this by developing tight lower bounds based on assignment costs which are embedded into a vector space for index-based acceleration.

\begin{table}[t]\centering
	\caption{Overview of methods for similarity search in graph databases.}%
	\label{table:relatedWork}%
	\setlength{\tabcolsep}{1pt}
	\resizebox{\textwidth}{!}{
		\begin{tabular}{lcrp{0.5\textwidth}ccc}
			\toprule
			\textbf{Name}&\textbf{Ref.}& \textbf{Year}& \textbf{Description} & \textbf{Bounds}& \textbf{Index} & \textbf{Exact} \\
			\midrule
			\textit{CStar} & \cite{2_ComparingStars}& 2009 & Optimal assignments on star structures & lower/upper& no & yes\\
			\textit{GSim} & \cite{32_GSimJoin}& 2012 & Approximation using path-based $q$-grams &lower&yes & yes\\
			\textit{Segos} & \cite{31_SEGOS} & 2012 & Two-level inverted index &lower/upper& yes & yes\\
			\textit{k-AT} & \cite{12_efficient} & 2012 & Index over tree-based $q$-grams &lower&yes& yes\\
			\textit{Pars} & \cite{3_partitionSim}& 2013 & Dynamic partitioning of graphs &lower & yes & yes\\
			\textit{Mixed} & \cite{ZhengZLWZ15}& 2015 & Partitioning and $q$-gram based filtering & lower & yes & yes \\
			\textit{MLIndex} & \cite{5_simMultiIndex}& 2017 & Partitioning graphs in a multi-layered Index  &lower& yes & yes\\
			\textit{Inves} & \cite{21_inves2019}& 2019& Incremental partitioning used in verification&lower/upper  & no & yes\\
			\textit{BSS\_GED} & \cite{CHEN2019762}& 2019& Filtering and efficient verification algorithm & lower & no & yes\\
			\textit{GHashing} & \cite{QinBS20} & 2020& Approximate filter via hashing and GNNs & --- & yes & no\\
			\textit{\textbf{EmbAssi}}&   & 2021& Embedding assignment costs & lower & yes & yes \\
			\bottomrule
	\end{tabular}}
	\vspace{-1\abovecaptionskip} %
\end{table}

\paragraph*{Our Contribution}
We develop multiple efficiently computable tight lower bounds for the graph edit distance, that allow exact filtering and can be used with an index for scalability to large databases.
Our techniques are shown to achieve a favorable trade-off between efficiency and effectivity in filtering.
Specifically, we make the following contributions.
\\\noindent
\textit{(1) Embedding Assignment Costs.} 
We build on a restricted version of the combinatorial assignment problem between two sets, where the ground costs for assigning individual elements are a tree metric.
With this constraint, the cost of an optimal assignment equals the $\ell_1$ distance between vectors derived from the sets and the weighted tree representing the metric~\cite{1_gedLinear}. We show that these vectors can be computed in linear time and optimized for combination with well-studied indices for vector data~\cite{DBLP:journals/corr/abs-1902-03616}. 
\\\noindent
\textit{(2) Lower Bounds.}
We formulate several assignment-based distance functions for graphs that are proven to be lower bounds on the graph edit distance.
We show that their ground cost functions are tree metrics and derive the corresponding trees, from which suitable vector representations are computed.
We propose bounds supporting uniform as well as non-uniform edit cost models for vertex labels.
Further bounds based on vertex degrees and labeled edges are introduced, some of which can be combined to obtain tighter lower bounds.
We analyze the proposed lower bounds and formally relate them to existing bounds from the literature.
\\\noindent
\textit{(3) EmbAssi.}
We use the vector representation for similarity search in graph databases following the filter-verification paradigm, building upon established indices for the Manhattan ($\ell_1$) distance on vectors.
Our approach supports range queries as well as $k$-nearest neighbor search using the optimal multi-step $k$-nearest neighbor search algorithm~\cite{25_optimalknn}.
This allows employing our approach in downstream machine learning and data mining methods such as nearest neighbors classification, local outlier detection~\cite{DBLP:journals/datamine/SchubertZK14}, or density-based clustering~\cite{EsterKSX96}.
\\\noindent
\textit{(4) Experimental Evaluation.}
We show that, while the proposed bounds are often close to or even outperform state-of-the-art bounds~\cite{29_ged_heuristics,2_ComparingStars}, they can be computed much more efficiently.
In the filter-verification framework, our approach obtains manageable candidate sets for verification in a very short time even in databases with millions of graphs, for which most competitors fail.
Our approach supports efficient construction of an index used for all query thresholds and is, compared to several competitors~\cite{5_simMultiIndex,32_GSimJoin}, not restricted to connected graphs with a certain minimum size.
We show that our approach can be combined with more expensive lower and upper bounds in a subsequent step to further reduce overall query time.

\section{Related Work}
\label{sec:relatedwork}
We summarize the related work on similarity search in graph databases and graph edit distance computation and conclude by a discussion motivating our approach.

\subsection{Similarity Search in Graph Databases}
Several methods for accelerating similarity search in graph databases have been proposed, see Table~\ref{table:relatedWork}.
Most approaches follow the filter-verification paradigm and rely on lower and upper bounds of the graph edit distance. These techniques focus almost exclusively on range queries and assume a uniform cost function for graph edit operations.
Most of the methods suitable for similarity search can be divided into two categories depending on whether they compare overlapping or non-overlapping substructures. 

Representatives of the first category are \textit{k-AT}~\cite{12_efficient}, \textit{CStar}~\cite{2_ComparingStars}, \textit{Segos}~\cite{31_SEGOS} and \textit{GSim}~\cite{32_GSimJoin}.
These methods are inspired by the concept of \textit{$q$-grams} commonly used for string matching.
In \cite{12_efficient} tree-based $q$-grams on graphs were proposed.
The \emph{$k$-adjacent tree} of a vertex $v \in V(G)$, denoted $k$-AT$(v)$, is defined as the top-$k$ level subtree of a breadth-first search tree in $G$, starting with vertex $v$.
For example, the $1$-AT$(v)$ is a tree rooted at $v$ with the neighbors of $v$ as children. These trees can be generated for each vertex of a graph and the graph can then be represented as the set of its $k$-ATs. Lower bounds for filtering are computed from these representations, which are organized in an inverted index.
\textit{CStar}~\cite{2_ComparingStars} is a method for computing an upper and lower bound on the graph edit distance using so-called \textit{star representations} of graphs, which consist of a $1$-AT for each vertex, called \emph{star}. %
An optimal assignment between the star representations of graphs regarding a ground cost function on stars the assignment cost yields a lower bound~\cite{2_ComparingStars}.
An upper bound can be obtained by using the cost of an edit path induced by the optimal assignment. 
\textit{Segos}~\cite{31_SEGOS} also uses these stars as (overlapping) substructures, but enhances the computation of the mapping distance and makes use of a two-layered index for range queries.
Another view on $q$-grams is given by the \textit{GSim} method~\cite{32_GSimJoin}, which uses path-based $q$-grams, i.e., simple path of length~$q$, instead of stars.
Since the number of path-based $q$-grams affected by an edit operation is lower than the number of tree-based $q$-grams, the derived lower bound is tighter~\cite{32_GSimJoin}.

The second category includes \textit{Pars}~\cite{3_partitionSim}, \textit{MLIndex}~\cite{5_simMultiIndex} and \textit{Inves}~\cite{21_inves2019}, which partition the graphs into non-overlapping substructures. They essentially obtain lower bounds based on the observation, that if $x$ partitions of a database graph are not contained in the query graph, the graph edit distance is at least $x$.
\textit{Pars} uses a dynamic partitioning approach to exploit this, while \textit{MLIndex} uses a multi-layered index to manage multiple partitionings for each graph.
\textit{Inves} is a method used to verify whether the graph edit distance of two graphs is below a specified threshold by first trying to generate enough mismatching partitions.
\textit{Mixed}~\cite{ZhengZLWZ15} combines the idea of $q$-grams and graph partitioning. 
First, a lower bound that uses the same idea as \textit{Inves}~\cite{21_inves2019}, but a different approach to find mismatching partitions, is proposed. Another lower bound based on so-called branch structures
(a vertex and its adjacent edges without the opposite vertex) is combined with the first one to gain an even tighter lower bound. This bound can be generalized to non-uniform edit costs and is referred to as \textit{Branch}~\cite{29_ged_heuristics}.
Recently, it has been proven that this bound is metric and its combination with an index to speed up similarity search for attributed graphs has been proposed~\cite{DBLP:conf/sisap/BauseBSK21}.

\subsubsection{Pairwise Computation of the Graph Edit Distance}
In the verification step, the remaining candidates have to be validated by computing the exact graph edit distance. Both general-purpose algorithms~\cite{34_blp} as well as approaches tailored to the verification step have been proposed~\cite{Chang2020}, which are usually based on depth- or breadth-first search~\cite{Gouda2016,Chang2020} or integer linear programming~\cite{34_blp}. 

On large graphs, these methods are not feasible and approximations are used~\cite{35_beamS,CHEN2019762,27_Riesen,1_gedLinear}. These can be obtained from the exact approaches, e.g., using beam search~\cite{35_beamS} or linear programming relaxations~\cite{29_ged_heuristics}. 
\textit{BeamD}~\cite{35_beamS} finds a sub-optimal edit path following the $A^*$ algorithm by extending only a fixed number of partial solutions.
A state-of-the-art approach is \textit{BSS\_GED}~\cite{CHEN2019762}, which reduces the search space based on beam stack search. It is not only used for computation of the exact graph edit distance, but also for similarity search by filtering with lower bounds during a linear database scan.
Recently, an approach using neural networks to improve the performance of the beam search algorithm was proposed~\cite{9458863}.

A successful technique referred to as \emph{bipartite graph matching}~\cite{27_Riesen} obtains a sub-optimal edit path from the solution of an optimal assignment between the vertices where the ground costs also encode the local edge structure.
The assignment problem is solved in cubic time using Hungarian-type algorithms~\cite{Burkard2012,Munkres} or in quadratic time using simple greedy strategies~\cite{Riesen2015}.
The running time was further reduced by defining ground costs for the assignment problem that are a tree metric~\cite{1_gedLinear}. This allows computing an optimal assignment in linear time by associating elements to the nodes of the tree and matching them in a bottom-up fashion. A tree metric gained from the Weisfeiler-Lehman refinement showed promising results. %

\subsubsection{Discussion}
Various upper and lower bounds for the graph edit distance are known, some of which have been proposed for similarity search, while others are derived from algorithms for pairwise computation and are not directly suitable for fast searching in databases.
Recently, an extensive study~\cite{29_ged_heuristics} of different bounds confirmed, that there is a trade-off between computational efficiency and tightness. Lower bounds based on linear programming relaxations and the linear assignment problem were found to be most effective. However, the computation of such bounds requires solving an optimization problem and the combination with indices is non-trivial.
Therefore, it has been proposed to compute graph embeddings optimized by graph neural networks, which reflect the graph edit distance, to make efficient index-based filtering possible~\cite{QinBS20}.
This and numerous other approaches~\cite{Li2019GraphMN,simgnn} that use neural networks to approximate the similarity of graphs do not compute lower or upper bounds on the graph edit distance and hence cannot be used to obtain exact results. Because of this, they are only suitable in situations in which incomplete answer sets are acceptable and are not in direct competition with exact approaches.

Recently, distance measures based on optimal assignments or, more generally, optimal transport (a.k.a.~Wasserstein distance) have become increasingly popular for structured data. A method for approximate nearest neighbor search regarding the Wasserstein distance has been proposed recently~\cite{30_scalableNNS}.
Another line of work studies special cases, which allow vector space embeddings, e.g., in the domain of kernels for structured data~\cite{KriegeGW16,wasser,1_gedLinear}. On that basis we develop embeddings of novel assignment-based lower bounds for the graph edit distance, which are effective and allow index-accelerated similarity search, while guaranteeing exact results.

\section{Preliminaries}
\label{sec:preliminaries}
We first give an overview of basic definitions concerning graph theory and database search. Then, we introduce tree metrics and the assignment problem, which play a major role in our new approach.

\subsection{Graph Theory}
\label{subsec:graphs}
A \emph{graph} $G = (V,E,\mu,\nu)$ consists of a set of vertices $V(G)=V$,
a set of edges $E(G)=E \subseteq V\times V$,
and labeling functions $\mu: V \rightarrow L$ and $\nu: E \rightarrow L$ for the vertices and edges. 
The labels $L$ can be arbitrarily defined.
We consider undirected graphs and denote an edge between $u$ and $v$ by $uv$.
The \emph{neighbors} of a vertex $v$ are denoted by $N(v) = \{u\mid  uv\in E(G)\}$ and the \emph{degree} of $v$ is $\delta(v) = |N(v)|$. The maximum degree of a graph $G$ is $\delta(G) = \max_{v\in V} \delta(v)$ and we let $\Delta = \max_{G\in \DB} \delta(G)$ for the graph dataset $\DB$.

A measure commonly used to describe the similarity of two graphs is the \textit{graph edit distance}.
An \textit{edit operation} can be deleting or inserting an isolated vertex or an edge or relabeling any of the two.
An \textit{edit path} between graphs $G$ and $H$ is a sequence $(e_1,e_2,\dots,e_k)$ of edit operations that transforms $G$ into~$H$. This means, that if we apply all operations in the edit path to $G$, we get a graph $G'$ that is isomorphic to $H$, i.e., we can find a bijection $\xi\colon V(G') \rightarrow V(H)$, so that $\forall v \in V(G'). \mu(v) = \mu(\xi(v)) \wedge \forall uv \in E(G'). \nu(uv) = \nu(\xi(u)\xi(v))$.
The graph edit distance is the cost of the (not necessarily unique) cheapest edit path.

\begin{definition}[Graph Edit Distance~\cite{27_Riesen}]
	Let $c$ be a function assigning non-negative costs to edit operations.
	The \emph{graph edit distance} between two graphs $G$ and $H$ is defined as
	$$d_{\textnormal{ged}}(G,H) = \min\left\{ \textstyle\sum\nolimits_{i=1}^{k} c(e_i) \,\mid \, (e_1, \dots, e_k) \in \Upsilon(G,H) \right\},$$ where $\Upsilon(G,H)$ denotes all possible edit paths from $G$ to $H$.
\end{definition}

Computation of the graph edit distance is \cNP-hard~\cite{2_ComparingStars}. Hence, exact computation is possible only for small graphs. There are several heuristics, see Section~\ref{sec:relatedwork}, many of which are based on solving an assignment problem.

\subsection{Optimal Assignments and Tree Metrics}
\label{subsec:treemetrics}
The assignment problem is a well-studied combinatorial optimization problem~\cite{Munkres,Burkard2012}.

\begin{definition}[Assignment Problem]
	\label{defAssProb}
	Let $A$ and $B$ be two sets with $\abs{A}=\abs{B}=n$ and $c\colon A \times B \to \mathbb{R}$ a ground cost function. 
	An \emph{assignment} between $A$ and $B$ is a bijection $f\colon A \to B$. The \emph{cost} of an assignment $f$ is $c(f) = \sum_{a\in A} c(a,f(a))$.
	The \emph{assignment problem} is to find an assignment with minimum cost.
\end{definition}
For an assignment instance $(A,B,c)$, we denote the cost of an optimal assignment by $d^c_\mathrm{oa}(A,B)$.
The assignment problem can be solved in cubic running time using a suitable implementation of the Hungarian method~\cite{Munkres,Burkard2012}. The running time can be improved when the cost function is restricted, e.g., to integral values from a bounded range~\cite{Duan2012}.
Of particular interest for our work is the requirement that the cost function is a tree metric, which allows to solve the assignment problem in linear time~\cite{1_gedLinear} and relates the optimal cost to the Manhattan distance, see Section~\ref{eac} for details.
We summarize the concepts related to these distances.

\begin{definition}[Metric] \label{metric}
	A \textit{metric} $d$ on $X$ is a function $d\colon X \times X \to \mathbb{R}$ that satisfies the following properties for all $x,y,z \in X$:
	(1)~$d(x,y)\geq 0$ (non-negativity), %
	(2)~$d(x,y)=0 \Longleftrightarrow x = y$ (identity of indiscernibles),
	(3)~$d(x,y)=d(y,x)$ (symmetry),
	(4)~$d(x,y)\leq d(x,z) + d(y,z)$ (triangle inequality). %
\end{definition}

The \emph{Manhattan metric} (also \emph{city-block} or \emph{$\ell_1$ distance}) is the metric function %
$d_m(\bm{x},\bm{y})= \norm{\smash{\bm{x}-\bm{y}}}_1 = \sum_{i=1}^{n} \mid x_i -y_i\mid$.
A \textit{tree} $T$ is an acyclic, connected graph.
To avoid confusion, we will call its vertices nodes.
A tree with non-negative %
edge weights $w: E(T) \rightarrow \mathbb{R}_{\geq 0}$ yields a function $d_{T\!,w}(u,v) = \sum_{e \in P(u,v)} w(e)$ on $V(T)$, where $P(u,v)$ is the unique simple path from $u$ to $v$ in $T$. 
\begin{definition}[Tree Metric]\label{defTreeMetric}
	A function $d: X \times X \to \mathbb{R}$ is a \emph{tree metric} if there is a tree $T$ with $X \subseteq V(T)$ and 
	strictly positive real-valued edge weights $w$,
	such that $d(u,v) = d_{T\!,w}(u,v)$, for all $u,v \in X$.
\end{definition}
Vice versa, every tree with strictly positive weights induces a tree metric on its nodes.
Equivalently, a metric $d$ is a tree metric iff $\forall v,w,x,y \in X.$ $ d(x,y)+d(v,w)\leq \max\{d(x,v)+d(y,w), d(x,w)+d(y,v)\}$~\cite{Semple2003}.
For such a tree with leaves $X$, a distinguished root, and the additional constraint that all paths from the root to a leaf have the same weighted length, the induced tree metric is an \emph{ultrametric}. Equivalently, a metric $d$ on $X$ is an ultrametric if it satisfies the strong triangle inequality $\forall x,y,z \in X.$ $ d(x,y) \leq \max\{d(x,z),d(y,z)\}$~\cite{Semple2003}.
In the following, we also allow edge weight zero. Therefore, the distances induced by a tree may violate property (2) of Definition~\ref{metric} and are therefore \emph{pseudometrics}. For the sake of simplicity, we still use the terms tree metric and ultrametric.

We consider the assignment problem, where the cost function $c\colon A \times B \to \mathbb{R}$ is a tree metric $d: X \times X \to \mathbb{R}$. To formalize the link between these two functions and, hence, Definitions~\ref{defAssProb} and \ref{defTreeMetric}, we introduce the map $\varrho\colon A \cup B \to X$.
Given a tree metric specified by the tree $T$ with weights $w$ and the map $\varrho$, the cost for assigning an object $a \in A$ to an object $b \in B$ is defined as $c(a,b)=d_{T\!,w}(\varrho(a), \varrho(b))$.
Note that $\varrho$ is not required to be injective. Therefore, $c$ may be a pseudometric even for trees with strictly positive weights. The input size of the assignment problem according to Definition~\ref{defAssProb} typically is quadratic in $n$ as $c$ is given by an $n\times n$ matrix. If $c$ is a tree metric, it can be compactly represented by the tree $T$ with weights $w$ having a total size linear in $n$.

\subsection{Searching in Databases}
\label{subsec:database}
Databases can store data in order to retrieve, insert or change it efficiently.
Regarding data analysis, retrieval (search) is usually the crucial operation on databases,
because it will be performed much more often than updates.
We focus on two types of similarity queries when searching a database $\DB$,
the first of which is the \textit{range query} for a radius $r$.

\begin{definition}[Range Query]
	Given a query object $q$ and a threshold $r$, %
	determine $\range{q}{r}= \{o \in \DB \mid d(o,q)\leq r\}$.
\end{definition}

A range query finds all objects with a distance no more than the specified range threshold $r$ to the query object $q$. 
If the distance $d$ is expensive to compute, it makes sense to use the so-called filter-verification principle.
In this approach different lower and upper bounds are used to filter out a hopefully large portion of the database. 
A function $d'$ is a \emph{lower bound} on $d$ if $d'(x,y)\leq d(x,y)$, and an \emph{upper bound} if $d'(x,y)\geq d(x,y)$ for all $x, y \in X$.
Clearly, objects where one of the lower bounds is greater than $r$ can be dismissed since the exact distance would be even greater.
Objects, where an upper bound is at most $r$ can be added to the result immediately.
Only the remaining objects need to be verified by computing the exact distance.

The second type of query considered here is the \textit{nearest neighbor query}, which returns the objects that are closest to the query object.

\begin{definition}[$k$-Nearest Neighbor Query, $k$nn Query]
	Given a query object $q$ and a parameter $k$, determine the smallest set $\kNN{q}{k}\subseteq \DB$, so that
	$\abs{\kNN{q}{k}}\geq k$ and $\forall o \in \kNN{q}{k}, \forall o^\prime\in \DB \setminus \kNN{q}{k}:\enskip d(o,q)< d(o^\prime,q).$
\end{definition}

In conjunction with range queries, it is preferable to return all the objects 
with a distance, that does not exceed the distance to the $k$th neighbor, which may be more than $k$ objects when tied.
That yields an equivalency of the results of $k$nn queries and range queries,
i.e., we have
$\range{q}{r}=\kNN{q}{\abs{\range{q}{r}}}$
and $\kNN{q}{k}=\range{q}{r_k}$,
where $r_k$ is the maximum distance in $\kNN{q}{k}$.

The optimal multi-step $k$-nearest neighbor search algorithm \cite{25_optimalknn} minimizes the number of candidates verified by using an incremental neighbor search, which returns the objects in ascending order regarding the lower bound.
As new candidates are discovered, their exact distance is computed. The current $k$th smallest exact distance is used as a bound
on the incremental search: once we have found at least $k$ objects with an exact distance smaller than the lower bound of all remaining objects, the result is complete. This approach is optimal in the sense that none of the exact distance computations could have been avoided~\cite{25_optimalknn}. 

\section{EmbAssi: Embedding Assignment Costs for Graph Edit Distance Lower Bounds}
\label{sec:ourMethod}

We define lower bounds for the graph edit distance obtained from optimal assignments regarding tree metric costs. These bounds are embedded in $\ell_1$ space and used for index-accelerated filtering.
In Section~\ref{eac} we describe the general technique for embedding optimal assignment costs for tree metric ground costs based on~\cite{1_gedLinear}.
In Section~\ref{efficient} we propose several embeddable lower bounds for the graph edit distance derived from such assignment problems. These are suitable for graphs with discrete labels and uniform edit costs and are generalized to non-uniform edit costs. %
In Sections~\ref{construct} and~\ref{query} we show how to use these bounds for both range and $k$-nearest neighbor queries and discuss optimization details.

\subsection{Embedding Assignment Costs}
\label{eac}%
Let $(A,B,c)$ be an assignment problem, where the cost function $c$ is a tree metric defined by the tree $T$ with weights $w$. The cost of an optimal assignment is equal to the Manhattan distance between vectors derived from the sets $A$ and $B$ using the tree $T$ and weights $w$~\cite{1_gedLinear}.
Recall that the cost of an assignment is the sum of the costs of all matched pairs.
A matched pair $(a,b)$ contributes the cost defined by the weight of the edges on the unique path between the nodes $\varrho(a)$ and $\varrho(b)$ in $T$. Hence, the total cost can be obtained from the number of times the edges occur on such paths.

\begin{figure}[t]
	\begin{subfigure}[b]{0.4\textwidth}
		\centering
		\includegraphics[width=0.6\textwidth]{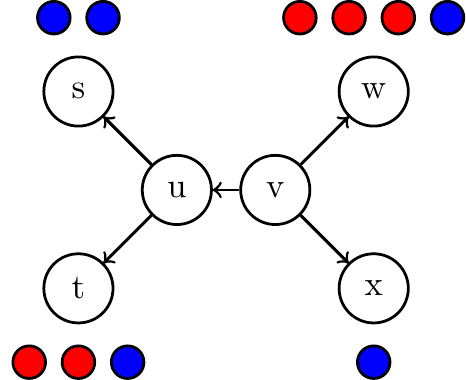}
		\subcaption{Tree $T$ representing metric $c$}
		\label{fig:embedassign:tree}
	\end{subfigure}
	\hfill
	\begin{subfigure}[b]{0.6\textwidth}\centering
		\setlength{\tabcolsep}{5pt}
		\renewcommand{\arraystretch}{1.3}
		\begin{tabular}[h]{c|rrrrr}
			& $\overleftarrow{su}$ & $\overleftarrow{tu}$ & $\overleftarrow{uv}$ & $\overleftarrow{wv}$ &$\overleftarrow{xv}$ \\
			\hline
			$\Phi_c(A)$ & 0 & 2 &2&3&0\\
			
			$\Phi_c(B)$
			& 2 & 1 &3&1&1\\
		\end{tabular}\vspace{1em}
		\subcaption{Embeddings}
		\label{fig:embedassign:embedding}
	\end{subfigure}
	\caption[An instance of the assignment problem and a tree metric]{ (\subref{fig:embedassign:tree}) An assignment problem $(A,B, c)$ with a tree $T$ representing the metric $c$. The elements of $A$ are denoted by {\large\textcolor{red}{$\bullet$}} and the elements of $B$ by {\large\textcolor{blue}{$\bullet$}} and associated to the nodes of $T$ by $\varrho$ as depicted. All edges have weight $1$.
		(\subref{fig:embedassign:embedding}) Embedding of $A$ and $B$ regarding $T$.
		The entry $\overleftarrow{uv}$ for the set $B$ counts the total number of {\large\textcolor{blue}{$\bullet$}} elements associated with the nodes $s$, $t$ and $u$ as indicated by the direction of the edge $uv$ in the tree. The assignment cost is $d^c_\mathrm{oa}(A,B) = 7$.}
	\label{fig:optass}
	\vspace{-1\abovecaptionskip}
\end{figure}

Let $S_{\overleftarrow{uv}}$ correspond to the number of elements of a set $S$, that are associated by the mapping~$\varrho$ with nodes in the subtree of $T$ containing~$u$, when the edge $uv$ is deleted (see Figure~\ref{fig:optass}).
It was shown in~\cite{1_gedLinear} that an optimal assignment has cost
\begin{equation}\label{eq:opt}
d^c_\mathrm{oa}(A,B)  = \sum\nolimits_{uv \in E(T)} \mid A_{\overleftarrow{uv}} - B_{\overleftarrow{uv}}\mid \cdot w(uv).
\end{equation}

Note that the roles of $u$ and $v$ are interchangeable and we indicate the choice by directing the edge accordingly.
Although this does not affect the assignment cost when applied consistently, there are subtle technical consequences, which we discuss for concrete tree metrics in Section~\ref{construct}.
Using $T$ and $w$ we can map sets to vectors having a component for every edge of $T$ defined as
$\Phi_c(S) = \left[\smash{S_{\overleftarrow{uv}}} \cdot w(uv)\right]_{uv \in E(T)}.$
From Eq.~\ref{eq:opt} it directly follows that the optimal assignment cost are
$$d^c_\mathrm{oa}(A,B) = \norm{\Phi_c(A) - \Phi_c(B)}_1.$$

This embedding of the optimal assignment cost into $\ell_1$ space is used in the following to obtain assignment-based lower bounds on the graph edit distance.

\subsection{Embeddable Lower Bounds}
\label{efficient}
Several lower bounds on the graph edit distance can be obtained from optimal assignments~\cite{29_ged_heuristics}.
However, these typically do not use a tree metric cost function, which complicates the embedding of assignment costs.
In \cite{1_gedLinear} two tree metrics, one based on Weisfeiler-Lehman refinement for graphs with discrete labels and one using clustering for attributed graphs, were introduced.
These, however, both do not yield a lower bound. 
We develop new lower bounds on the graph edit distance from optimal assignment instances, which have a tree metric ground cost function.
Most similarity search techniques for the graph edit distance assume a uniform cost model, where every edit operation has the same cost. We also support variable cost functions and discuss choices that are supported by our approach. We use $c_v$/$c_e$ to denote the costs of inserting or deleting vertices/edges and $c_{vl}$/$c_{el}$ for the costs of changing the respective label.

\subsubsection{Vertex Label Lower Bounds}\label{sec:vllb}
A natural method for defining a lower bound on the graph edit distance is to just take the labels into account ignoring the graph structure.
We first discuss the case of uniform cost for changing a label, which is common for discrete labels. Then, non-uniform costs are considered.

\paragraph{Uniform Cost Functions}
Clearly, each vertex of one graph, that cannot be assigned to a vertex of the other graph with the same label has to be either deleted or relabeled.
The idea leads to a particularly simple assignment instance when we assume fixed costs $c_{vl}$ and $c_v$.
Let $\Gone$ and $\Gtwo$ be two graphs with $n$ respectively $m$ vertices. Following the common approach to obtain an assignment instance~\cite{27_Riesen}, we extend $\Gone$ by $m$ and $\Gtwo$ by $n$ dummy nodes denoted by $\epsilon$.\footnote{One can also add only a single dummy node and modify the definition of an assignment~\cite{29_ged_heuristics}. However, this makes no difference for our technique, which embeds the assignment costs. We discuss how dummy vertices can be avoided in Section~\ref{construct}.}
We consider the following assignment problem.

\begin{definition}[Label Assignment]
	The \emph{label assignment} instance for $\Gone$ and $\Gtwo$ is given by $(V(\Gone), V(\Gtwo), c_{\text{llb}})$, where the ground cost function is
	$$c_{\text{llb}}(u,v) = \begin{cases}
	0    & \text{ if } \mu(u) = \mu(v) \text{ or } u=v=\epsilon \\
	c_{vl} & \text{ if } \mu(u) \neq \mu(v) \\
	c_v  & \text{ if either } u=\epsilon \text{ or } v=\epsilon.
	\end{cases}$$
\end{definition}
We define $\LLB(\Gone,\Gtwo)=d^{c_{\text{llb}}}_\mathrm{oa}(V(\Gone),V(\Gtwo))$ and show that it provides a lower bound on the graph edit distance.
\begin{proposition}[Label lower bound]\label{llb_lower_bound}
	For any two graphs $\Gone$ and $\Gtwo$, we have $\LLB(\Gone,\Gtwo) \leq \GED(\Gone, \Gtwo)$.
	\begin{proof}
		Every assignment directly induces a set of edit operations, which can be arranged to form an edit path. Vice versa, every edit path can be represented by an assignment~\cite{27_Riesen,29_ged_heuristics}.
		Let $\bm{e}$ be a minimum cost edit path. We construct an assignment $f$ from the vertex operations in $\bm{e}$, where the deletion of $v$ is represented by $(v,\epsilon) \in f$, insertion by $(\epsilon, v) \in f$, and relabeling of the vertex $u$ with the label of $v$ by $(u,v) \in f$, where $u,v \neq \epsilon$.
		We have $c(\bm{e}) = Z_v+Z_e$, where $Z_v$ and $Z_e$ are the costs of vertex and edge edit operations, respectively. According to the definition of $c_{\text{llb}}$ and the construction of $f$ we have $Z_v=c_{\text{llb}}(f)$. 
		An optimal assignment $o$ satisfies $c_{\text{llb}}(o) \leq c_{\text{llb}}(f)$ and $\LLB(\Gone,\Gtwo)=c_{\text{llb}}(o) \leq c_{\text{llb}}(f) \leq c(\bm{e})=\GED(\Gone, \Gtwo)$ follows, since  $Z_e \geq 0$. 
	\end{proof}
\end{proposition}

To obtain embeddings, we investigate for which choices of edit costs the ground cost function $c_{\text{llb}}$ is a tree metric.

\begin{figure}
	\centering
	\begin{subfigure}[b]{0.2\columnwidth}
		\centering
		$\Gone$ \includegraphics[scale=0.7]{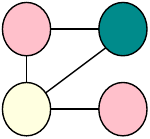} \\
		$\Gtwo$ \includegraphics[scale=0.7]{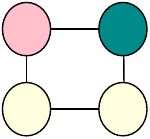}
		\caption{Graphs}
		\label{fig:llb:graphs}
	\end{subfigure}
	\hfill
	\begin{subfigure}[b]{0.3\columnwidth}
		\centering
		\includegraphics[scale=.45]{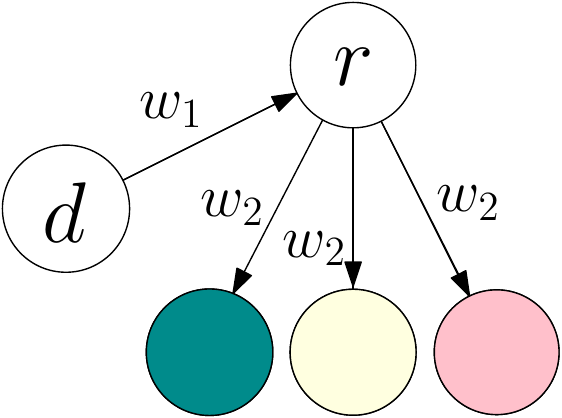}
		\caption{Weighted tree}
		\label{fig:llb:tree}
	\end{subfigure}
	\hfill
	\begin{subfigure}[b]{0.4\columnwidth}
		\centering\small
		\begin{tabular}[c]{|r|}
			\hline
			$\Phi (\Gone)$\\
			\hline
			$4 \cdot w_1$\\
			$1 \cdot w_2$\\
			$1 \cdot w_2$\\
			$2 \cdot w_2$\\
			\hline
		\end{tabular}
		\begin{tabular}[c]{|r|}
			\hline
			$\Phi (\Gtwo)$ \\
			\hline
			$4 \cdot w_1$\\
			$1 \cdot w_2$\\
			$2 \cdot w_2$\\
			$1 \cdot w_2$\\
			\hline
		\end{tabular}
		\caption{Embeddings}
		\label{fig:llb:embed}
	\end{subfigure}
	\caption{Two graphs $\Gone$ and $\Gtwo$~(\subref{fig:llb:graphs}), the weighted tree representing the ground cost function $c_{\text{llb}}$~(\subref{fig:llb:tree}), and the derived embeddings~(\subref{fig:llb:embed}). The weights are $w_1 = c_v-\tfrac12 c_{vl}$ and $w_2=\tfrac12 c_{vl}$. The entries of the vectors correspond to the edges of the tree, from left to right, arrows indicate the direction used when counting elements.
	}%
	\label{fig:llb}
	\vspace{-1\abovecaptionskip} %
\end{figure}

\begin{proposition}[LLB tree metric]\label{llb:tree_metric}
	The ground cost function $c_{\text{llb}}$ is a tree metric if and only if $c_{vl} \leq 2c_v$.
	\begin{proof}
		First we assume $c_{vl} \leq 2c_v$ and define a tree $T$ with a central node $r$ having a neighbor for every label $l \in L$ and a neighbor~$d$. Let $w(rl)=\tfrac12 c_{vl}$ for all $l \in L$ and $w(rd)=c_v-\tfrac12 c_{vl}$, cf.~Fig.~\ref{fig:llb:tree}. The assumption guarantees that all weights are non-negative. We consider the map $\varrho(v)=\mu(v)$ for $v\neq \epsilon$ and $\varrho(\epsilon)=d$. We observe that $c_{\text{llb}}(u,v)=d_{T\!,w}(\varrho(u), \varrho(v))$ by verifying the three cases.
		
		The reverse direction is proven by contradiction. Assume $c_{vl} > 2c_v$ and $c_{\text{llb}}$ a tree metric. Let $u$ and $v$ be two vertices with $\mu(u)\neq\mu(v)$, then $c_{\text{llb}}(u,v) =c_{vl}$ and $c_{\text{llb}}(u, \epsilon)=c_{\text{llb}}(\epsilon, v) = c_v$. Therefore,
		$c_{\text{llb}}(u,v) > c_{\text{llb}}(u, \epsilon)+c_{\text{llb}}(\epsilon, v)$ contradicting the triangle inequality, Definition~\ref{metric},~(4). Thus, $c_{\text{llb}}$ is not a metric and, in particular, not a tree metric contradicting the assumption.
	\end{proof}
\end{proposition}

The requirement $c_{vl} \leq 2c_v$ states that relabeling a vertex is at most as expensive as deleting and inserting it with the correct label. This is generally reasonable and not a severe limitation.
Because the proof is constructive, it allows us to represent $c_{\text{llb}}$ by a weighted tree,
from which we can compute the graph embedding representing the assignment costs following the approach described in Section~\ref{eac}.

Fig.~\ref{fig:llb} illustrates the embedding of the label lower bound for an example. The tree representing the cost function is shown in Fig.~\ref{fig:llb:tree}.
The weight of the edge from the dummy node to the root is chosen, such that the path length from a label to the dummy node is $c_v$. 
Fig.~\ref{fig:llb:embed} shows the vectors $\Phi$ of the two example graphs, which allows obtaining $\LLB(\Gone,\Gtwo)=\norm{\Phi (\Gone) - \Phi (\Gtwo)}_1 = c_{vl}$ as the Manhattan distance.

\paragraph{Non-Uniform Cost Functions}
We have discussed the case where changing one label into another has a fixed cost of $c_{vl}$. In general, the cost for this may depend on the two labels involved, i.e., we assume that a cost function $c_{vl}\colon L\times L \to \mathbb{R}_{\geq0}$ is given.
Two common scenarios can be distinguished: 
First, $L$ is a (small) finite set of labels that are similar to varying degrees. An example are molecular graphs, where the costs are defined based on vertex labels encoding their pharmacophore properties~\cite{Hernandez2019}. Second, $L$ is infinite (or very large), e.g., vertices are annotated with coordinates in $\mathbb{R}^2$ and the cost is defined as the Euclidean distance. We propose a general method and then discuss its applicability to both scenarios.

We can extend the tree defining the metric used in the above paragraph to allow for more fine-grained vertex relabel costs. To this end, an arbitrary ultrametric tree on the labels $L$ is defined, where the node $d$ representing deletions is added to its root $r$. Recall that in an ultrametric tree the lengths of all paths from the root to a leaf are equal to, say, $u$. We define the weight of the edge between $r$ and $d$ as $c_v - u$ and observe that $c_v \geq u$ is required to obtain a valid tree metric in analogy to the proof of Proposition~\ref{llb:tree_metric}. 

To obtain an ultrametric tree that reflects the given edit cost function $c_{vl}$, we employ hierarchical clustering. 
To guarantee that the assignment costs are a lower bound on the graph edit distance, it is crucial that interpreting the hierarchy as an ultrametric tree will underestimate the real edit costs. For optimal results, we would like to obtain a tight lower bound.
We formalize the requirements. Let $c_{vl}\colon L{\times} L \to \mathbb{R}_{\geq0}$ be the given cost function and $d_{hc}\colon L{\times} L \to \mathbb{R}_{\geq0}$ the ultrametric induced by hierarchical clustering of $L$ with cost function $c_{vl}$. 
Let $U^{-}\!(c_{vl})$ be the set of all ultrametrics that are lower bounds on $c_{vl}$. There is a unique ultrametric $d^* \in U^{-}\!(c_{vl})$ defined as $d^*(l_1,l_2) = \sup_{d \in U^{-}\!(c_{vl})} \{d(l_1,l_2)\}$ for all $l_1, l_2 \in L$ \cite{slc_bock}. This $d^*$ is an upper bound on all ultrametrics in $U^{-}\!(c_{vl})$, a lower bound on $c_{vl}$ and called the \emph{subdominant} ultrametric to~$c_{vl}$.
The subdominant ultrametric is generated by single-linkage hierarchical clustering~\cite{slc_bock}, which therefore is, in this respect, optimal for our purpose.
In particular, it reconstructs an ultrametric tree if the original costs are ultrametric. Moreover, single-linkage clustering can be implemented with running time $O(\abs{L}^2)$~\cite{slink}.

For a finite set of labels $L$, our method is a viable solution if the edit cost function $c_{vl}$ is close to an ultrametric and $L$ is small.
If $L$ is infinite, we need to approximate it with a finite set through quantization or space partitioning. The realization of such an approach preserving the lower bound property depends on the specific application and is hence not further explored here.

\subsubsection{Degree Lower Bound}
The $\LLB$ does not take the graph structure into account. We now introduce the degree lower bound, which focuses on how many edges have to be inserted or deleted at the minimum. 
When deleting or inserting vertices, all of the adjacent edges have to be deleted or inserted as well. If two vertices with differing degrees are assigned to one another, again edges have to be deleted or inserted accordingly.
As in Section~\ref{sec:vllb}, we extend the graphs $\Gone$ and $\Gtwo$ by dummy nodes $\epsilon$ and define an assignment problem.

\begin{definition}[Degree Assignment]	The \emph{degree assignment} instance for $\Gone$ and $\Gtwo$ is given by $(V(\Gone), V(\Gtwo), c_{\text{dlb}})$, where the ground cost function is	$c_{\text{dlb}}(u,v) = \tfrac12 c_e \mid\delta(u)-\delta(v)\mid$ with $\delta(\epsilon):=0$ for the dummy nodes.
\end{definition}
We define $\DLB(\Gone,\Gtwo)=d^{c_{\text{dlb}}}_\mathrm{oa}(V(\Gone),V(\Gtwo))$,
and show that it is a lower bound.
\begin{proposition}[Degree lower bound]\label{pr:dlb}
	For any two graphs $\Gone$ and $\Gtwo$, we have $\DLB(\Gone,\Gtwo) \leq \GED(\Gone, \Gtwo)$.
	\begin{proof}	Using the same arguments as in the proof of Proposition~\ref{llb:tree_metric}, let $\bm{e}$ be a minimum cost edit path and $f$ an assignment that induces $\bm{e}$. We divide the costs $c(\bm{e}) = Z_v{+}Z_e$ into costs $Z_v$ and $Z_e$ of vertex and edge edit operations. For the matched vertices $v$ and $f(v)$ at least $\mid\delta(v)-\delta(f(v))\mid$ edges must be deleted or inserted to balance the degrees; in case of insertion and deletion all adjacent edges must be inserted or deleted. Since each edge edit operation increases or decreases the degree of its two endpoints by one, the sum of these costs over all vertices must be divided by two and $c_{\text{dlb}}(f)=Z_e$ follows.
	\end{proof}
\end{proposition}

To obtain an embedding, we show that $c_{\text{dlb}}$ is a tree metric.

\begin{proposition}[DLB tree metric]\label{dlb:tree_metric}
	The ground cost function $c_{\text{dlb}}$ is a tree metric.
	\begin{proof}
		To prove that $c_{\text{dlb}}$ is a tree metric, we construct a tree $T$ with edge weights $w$ and a map $\varrho$, so that $c_{\text{dlb}}(u,v) = d_{T\!,w}(\varrho(u),\varrho(v))$.
		Let $T$ have nodes $V(T)=\{r=0, 1,\dots,\Delta\}$ %
		and edges $E(T)=\{ij \mid j=i{+}1\}$ with weight $w_1=\tfrac12 c_e$. Since $c_e$ cannot be negative, all edge weights are non-negative. We consider the map $$\varrho(u)=\begin{cases}
		r & \text{if } u=\epsilon \text{ or } \delta(u)=0 \\
		\delta(u)   & \text{otherwise.} \\
		\end{cases}$$
		It can easily be seen, that $c_{\text{dlb}}(u,v) = d_{T\!,w}(\varrho(u),\varrho(v))$ by verifying the path lengths in the tree.
	\end{proof}
\end{proposition}

\begin{figure}
	\centering
	\begin{subfigure}[b]{0.2\columnwidth}
		\centering
		$\Gone$ \includegraphics[scale=0.7]{g1} \\
		$\Gtwo$ \includegraphics[scale=0.7]{g2}
		\caption{Graphs}
		\label{fig:dlb:graphs}
	\end{subfigure}
	\hfill
	\begin{subfigure}[b]{0.3\columnwidth}
		\centering
		\includegraphics[scale=.4]{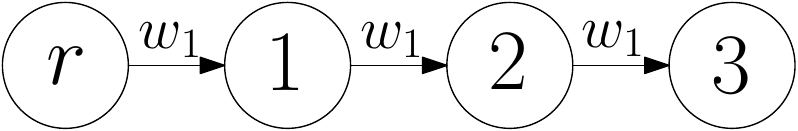}
		\caption{Weighted tree}
		\label{fig:dlb:tree}
	\end{subfigure}
	\hfill
	\begin{subfigure}[b]{0.4\columnwidth}
		\centering\small
		\begin{tabular}[c]{|r|}
			\hline
			$\Phi(\Gone)$ \\
			\hline
			$4 \cdot w_1$\\
			$3 \cdot w_1$\\
			$1 \cdot w_1$\\
			\hline
		\end{tabular}
		\begin{tabular}[c]{|r|}
			\hline
			$\Phi(\Gtwo)$ \\
			\hline
			$4 \cdot w_1$\\
			$4 \cdot w_1$\\
			$0 \cdot w_1$\\
			\hline
		\end{tabular}
		\caption{Embeddings}
		\label{fig:dlb:embed}
	\end{subfigure}
	\caption{Two graphs $\Gone$ and $\Gtwo$~(\subref{fig:dlb:graphs}), the weighted tree representing the cost function $c_{\text{dlb}}$~(\subref{fig:dlb:tree}), and the derived embeddings~(\subref{fig:dlb:embed}).
	}%
	\label{fig:dlb}
	\vspace{-1\abovecaptionskip} %
\end{figure}

The proof gives a concept to construct a tree representing the \textit{DLB} cost function. As there is no difference between a vertex with degree 0 and a dummy vertex, they can both be assigned to the root node $r$.
Note, that the edge labels are not taken into account by this lower bound and edge insertion and deletion are not distinguished.
Fig.~\ref{fig:dlb} illustrates the embedding of the degree lower bound, which yields $\DLB(\Gone,\Gtwo) = c_{e}$ for the running example. 

\subsubsection{Combined Lower Bound}\label{sec:clb}
We can combine \textit{LLB} and \textit{DLB} to improve the approximation.

\begin{definition}[CLB]
	The \emph{combined lower bound} between $\Gone$ and $\Gtwo$ is defined as
	$\CLB(\Gone,\Gtwo)= \LLB(\Gone,\Gtwo) + \DLB(\Gone,\Gtwo).$
\end{definition}
We show, that $\CLB$ is a lower bound on the graph edit distance. Note that this lower bound is based on the two assignments given by $\LLB$ and $\DLB$, which are not necessarily equal.

\begin{lemma}\label{lem:split_cost}
	Let $c_1, c_2$ and $c$ be ground cost functions on $X$ and $c(x,y)=c_1(x,y)+c_2(x,y)$ for all $x,y \in X$. 
	Then for any $A,B \subseteq X$, $\abs{A}=\abs{B}=n$, the inequality
	$d^{c_1}_\mathrm{oa}(A,B) + d^{c_2}_\mathrm{oa}(A,B) \leq d^{c}_\mathrm{oa}(A,B)$ holds. 
	\begin{proof}
		Let $o_1, o_2$ and $o$ be optimal assignments between $A$ and~$B$ regarding the ground costs $c_1, c_2$ and $c$, respectively. 
		Due to the optimality we have $c_1(o_1) \leq c_1(o)$ and $c_2(o_2) \leq c_2(o)$. Hence, 
		$d^{c_1}_\mathrm{oa}(A,B) + d^{c_2}_\mathrm{oa}(A,B) = c_1(o_1) + c_2(o_2)\leq c_1(o)+ c_2(o) = c(o) = d^{c}_\mathrm{oa}(A,B)$.
	\end{proof}
\end{lemma}

\begin{proposition}[Combined lower bound]
	For any two graphs $\Gone$ and $\Gtwo$, we have $\CLB(\Gone,\Gtwo) \leq \GED(\Gone, \Gtwo)$.
	\begin{proof}
		Let $\bm{e}$ be a minimum cost edit path and $f$ an assignment that induces $\bm{e}$. We divide the costs $c(\bm{e}) = Z_v+Z_e$ into costs $Z_v$ and $Z_e$ of vertex and edge edit operations. From the proof of Propositions~\ref{llb_lower_bound} and~\ref{pr:dlb} we know that $Z_v \geq c_{\text{llb}}(f)$ and $Z_e \geq c_{\text{dlb}}(f)$ and, hence, $c(\bm{e})\geq c_{\text{llb}}(f) + c_{\text{dlb}}(f)$. Application of Lemma~\ref{lem:split_cost} yields
		$\CLB(\Gone,\Gtwo) = c_{\text{llb}}(f_1)+c_{\text{dlb}}(f_2) \leq c_{\text{llb}}(f)+c_{\text{dlb}}(f) \leq c(\bm{e}) = \GED(\Gone, \Gtwo)$, where $f_1$ and $f_2$ are optimal assignments regarding $c_{\text{llb}}$ and $c_{\text{dlb}}$.
	\end{proof}
\end{proposition}
This lower bound is at least as tight as the ones it consists of and, therefore, most promising.
The combined lower bound is embedded by concatenating the vectors for \textit{LLB} and \textit{DLB}.

\subsection{Analysis}\label{analysis}
We provide a theoretical comparison of our proposed bounds to existing lower bounds and also give details on the time complexity of our approach.

\subsubsection{Comparison with Existing Bounds}\label{relation}
We relate the $\CLB$ to two well-known lower bounds when applied to graphs with vertex labels. 
The \emph{simple label filter} (\SLF) is the intersection of vertex and edge label multisets, i.e.,
$\SLF(G,H) = \abs{L_V(G) \cap L_V(H)} + \abs{\abs{ E(G)} -\abs{E(H)}}$ in our case, where $L_V$ denotes the vertex label multiset of a graph. Although simple, this bound is often found to be selective~\cite{21_inves2019} and, therefore, widely-used~\cite{32_GSimJoin,3_partitionSim}.
A very effective bound according to~\cite{29_ged_heuristics} is \textit{BranchLB} based on general optimal assignments. Several variants have been proposed~\cite{27_Riesen,ZhengZLWZ15,29_ged_heuristics} with at least cubic worst-case time complexity. In our case, \textit{BranchLB} is the cost of the optimal assignment regarding the ground costs $c_{\text{branch}}(u,v) = c_{\text{llb}}(u,v)+c_{\text{dlb}}(u,v)$. 
Note that $c_{\text{branch}}$ in general is not a tree metric. 
\SLF{} assumes $c_v=c_{vl}=c_e=1$ and we consider this setting although $\CLB$ and \textit{BranchLB} are more general.
Using counting arguments and Lemma~\ref{lem:split_cost} we obtain the following relation:
\begin{proposition}\label{pr:rel}
	For any two vertex-labeled graphs $\Gone$ and $\Gtwo$, $$\SLF(\Gone,\Gtwo)\,{\leq}\,\CLB(\Gone,\Gtwo)\,{\leq}\,\BranchLB(\Gone,\Gtwo)\,{\leq}\,\GED(\Gone, \Gtwo).$$
\end{proposition}

Experimentally we show in Section~\ref{sec:evaluation} that  our combined lower bound is close to \BranchLB{} for a wide-range of real-world datasets, but is computed several orders of magnitude faster and allows indexing. This makes it ideally  suitable for fast pre-filtering and search.

\subsubsection{Time Complexity}\label{time}
We first consider the time required for generating the vector $\Phi_c(S)$ for a set $S$ and tree $T$ defining the ground cost function $c$.
\begin{proposition}\label{runtime:tree}
	Given a set $S$ and a weighted tree $T$ representing the ground cost function $c$, the vector $\Phi_c(S)$ can be computed in $O(\abs{V(T)}+\abs{S})$ time.
	\begin{proof}
		We first associate the elements of $S$ with the nodes of $T$ via the map $\varrho$ and then traverse $T$ starting from the leaves progressively moving towards the center. The order guarantees that when the node $u$ is visited, exactly one of its neighbors, say $v$, has not yet been visited. Then $S_{\overleftarrow{uv}}$ can be obtained as $\sum_{w\in N(u)\setminus\{v\}} S_{\overleftarrow{wu}}$ from the values computed previously. The tree traversal and computation of $S_{\overleftarrow{uv}}$ for all $uv\in E(T)$ takes $O(\abs{V(T)})$ total time. Together with the time for processing the set $S$ we obtain $O(\abs{V(T)}+\abs{S})$ time.
	\end{proof}
\end{proposition}

The time complexity of the different bounds depends on the size of the tree representing the metric and the size of the graphs.
\begin{proposition}\label{runtime:lb}
	The bounds $c_{\text{llb}}$, $c_{\text{dlb}}$ and $c_{\text{clb}}$ for two graphs $G$ and $H$ can be computed in $O(\abs{V(G)}+\abs{V(H)})$ time.
	\begin{proof}
		First the tree $T$ defining the metric is computed. For the different tree metrics, the trees sizes are linear in the number of nodes of the two graphs $G$ and $H$: For $c_{\text{llb}}$ the tree (denoted $T_{\text{llb}}$) has size $\abs{L_v(G)\cup L_v(H)}+2$, where $L_v(G)$ denotes the set of vertex labels occurring in $G$. The tree consists of a node for each vertex label plus a dummy and a central node. In the worst case, where all labels occur only once, the tree is of size $\abs{V(G)}+\abs{V(H)}+2$.
		For $c_{\text{dlb}}$, we have $\abs{V(T_{\text{dlb}})} =\max(\delta(G),\delta(H))+1$, since there is a node for each vertex degree, up to the maximum degree (including degree $0$).
		As shown in Proposition~\ref{runtime:tree}, the vector $\Phi_c(G)$ can then be computed in time $O(\abs{V(T)}+\abs{V(G)})$. For $c_{\text{clb}}$ we concatenate the vectors of $c_{\text{llb}}$ and $c_{\text{dlb}}$.
		The Manhattan distance between two vectors is computed in time linear in the number of components, which is $O(\abs{V(T)})$.
		Thus, the total running time for computing any of the bounds is in $O(\abs{V(G)}+\abs{V(H)})$.
		
	\end{proof}
\end{proposition}

The bound $\SLF$ also has a linear time complexity while \textit{BranchLB} requires $O(n^2 \Delta^3 + n^3 )$ time for graphs with $n$ vertices and maximum degree $\Delta$~\cite{29_ged_heuristics}. Our new approach matches the running time of $\SLF$ but in most cases yields tighter bounds, cf.~Proposition~\ref{pr:rel} and our experimental evaluation in Section~\ref{sec:evaluation}. Hence, it provides a favorable trade-off between efficiency and quality and at the same time can conveniently be combined with indices.

\section{EmbAssi for Graph Similarity Search}
\label{sec:ourMethod2}
\begin{figure}[tb]
	\centering
	\includegraphics[width=0.9\linewidth]{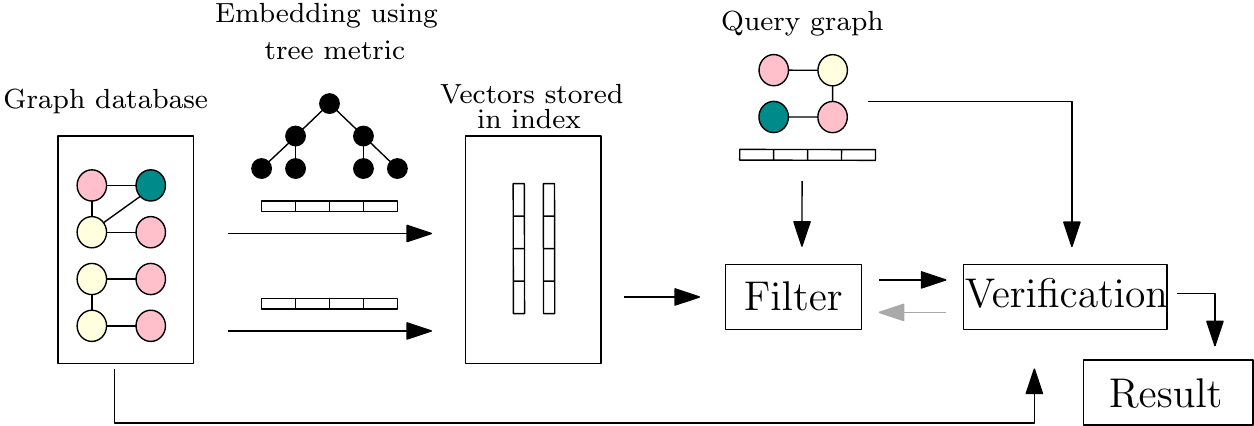}
	\caption{Overview of our pipeline for graph similarity search. In a preprocessing step the embeddings of all database graphs under the specified tree metric are computed and stored in an index. Then similarity search queries are answered by computing the embedding of the query graph and filtering regarding the Manhattan distance. In $k$-nearest neighbor search, the information gained from refining candidates is used to reduce the search range. This is indicated with a gray arrow. %
	}
	\label{fig:ubersicht}
\end{figure}

We use the proposed lower bounds for similarity search by computing embeddings for all the graphs in the database in a preprocessing step. Given a query graph, we compute its embedding and realize filtering utilizing indices regarding the Manhattan distance.
The approach is illustrated in Fig.~\ref{fig:ubersicht}.
Algorithm~\ref{alg:index} shows how the preprocessing is done, exemplary for $\LLB$: We construct the tree metric based on the labels and associate the vertices with the leaves of the tree. Then for each graph the embedding is computed using Algorithm~\ref{alg:vectors}.

Several technical details must be considered. The choice of how to direct the edges has a huge impact on the resulting vectors and of course using a suitable index is also important. In the following, we briefly discuss our choices and explain how similarity search queries can be answered.
\begin{algorithm}[tb]
	\caption{Construction for $\LLB$}\label{alg:index}
	\begin{algorithmic}
		\Procedure{constructIndex}{set of graphs $DB$}
		\State $T$ $\gets$ new tree with root $\epsilon$ and child $r$ 
		\State $w((\epsilon,r)) \gets c_v-\tfrac12 c_{vl}$
		\Comment Set edge weight
		\ForAll{$G \in DB$}
		\Comment Construct tree metric $\LLB$
		\ForAll{$v \in V(G)$}
		\If{$\mu(v) \notin T$}
		\State add node $\mu(v)$ as child of $r$ with $w((r,\mu(v)))=\tfrac12 c_{vl}$
		\State  $\varrho(v) \gets \mu(v)$
		\EndIf
		\EndFor
		\EndFor
		\ForAll{$G\in DB$}
		\Comment Compute embedding under tree metric
		\State $\Phi(G) \gets$ \Call{computeVector}{$G$, $T$}
		\EndFor
		\State Create index($\{\Phi(G) \mid G \in DB \}$, Manhattan distance)
		\EndProcedure
	\end{algorithmic}
\end{algorithm}
\begin{algorithm}[tb]
	\caption{Vector Computation}\label{alg:vectors}
	\begin{algorithmic}
		\Procedure{computeVector}{graph $G$, tree $T$}
		\State $\Phi(G) \gets$ sparse vector
		\State $S \gets$ relevant subtree$(V(G), T)$
		\Comment{$\forall v \in V(G):$ $\varrho(v)$ and path to root}
		\State $count[n] \gets 0, \forall n \in V(S)$ 
		\ForAll{$v \in V(G)$}
		\State $count[\varrho(v)] \gets count[\varrho(v)]+1$
		\Comment{Count vertices at leaves}
		\EndFor
		\State $leaves \gets$ leaves of $S$
		\While{$l \gets leaves$.dequeue() \textbf{and} $l \neq$ root of $S$}
		\State $\Phi(G)[(l,p)] \gets w((l,p)) \cdot count[l]$
		\State $count[p] = count[p] + count[l]$
		\Comment{Propagate vertex count to parent}
		\State delete $l$ from $S$
		\If{$\delta(p)\leq 1$}
		\Comment{$p$ is now a leaf itself}
		\State $leaves$.enqueue($p$)
		\EndIf
		\EndWhile
		\State \Return $\Phi(G)$
		\EndProcedure
	\end{algorithmic}
\end{algorithm}

\subsection{Index Construction}\label{construct}

We compute the vectors for all graphs in the database and store them in an index to accelerate queries.
When defining the bounds, we considered the pairwise comparison of two graphs and added dummy vertices to obtain graphs of the same size.
We have chosen the direction of edges in the trees representing the metrics carefully, cf.~Section~\ref{eac}, to generate consistent embeddings for the entire database. By rooting the trees at the node $\varrho(\epsilon)$ representing the dummy vertices (see Algorithm~\ref{alg:index}) and directing all edges towards the leaves the dummy vertices are not counted in any entry of the vectors, see Fig.~\ref{fig:llb}--\ref{fig:dlb}.
Moreover, this choice often leads to sparse vectors, e.g., for the \textit{LLB}, where every entry just counts the number of vertices with one specific label. Labels that only appear in a small fraction of the graphs in the database then lead to zero-entries in the vectors of the other graphs and sparse data structures become beneficial. Moreover, this simplifies to dynamically add new vertex labels without requiring to update all existing vectors in the database.
Using sparse vector representations, Algorithm~\ref{alg:vectors} can be implemented in time $O(\abs{V(G)})$ by considering only the relevant part of $T$. This is the subtree $S$ formed by the nodes of $T$ to which vertices of $G$ are assigned to via $\varrho$, and the nodes on the path to the root from these nodes. The subtree $S$ is processed in a bottom-up fashion computing a non-zero component of $\Phi(G)$ in each step. Note that $S$ can be maintained and modified with low overhead using flags to indicate whether a node of $T$ is contained in $S$.

The choice of a suitable index is crucial for the performance of our approach.
We chose to use the cover tree~\cite{BeygelzimerKL06} because our data is too high-dimensional for the popular k-d-tree, and our vectors have many zeros and discrete values.
The cover tree is a good choice for an in-memory index because of its lightweight construction, low memory requirements, and good metric pruning properties.
It is usually superior to the k-d-tree or R-tree if the data stored is high-dimensional but still has a small doubling dimension.

\subsection{Queries}\label{query}

For similarity search, we compute the embedding of the query graph and use the index for similarity search
regarding Manhattan distance. The index takes responsibility to disregard parts of the database that are too far away from the query object.
In $k$-nearest neighbor search, we use the optimal multi-step $k$-nearest neighbor search \cite{25_optimalknn} as described in Section~\ref{subsec:database} to stop the search as early as possible and compute the minimum necessary number of exact graph edit distances.
Our lower bounds are especially useful for this because it is well understood how to index data for ranking by Manhattan distance.
Further exact distance computations (in particular for range queries) can be avoided by checking additional bounds
similar to \textit{Inves}~\cite{21_inves2019} or \textit{BSS\_GED}~\cite{CHEN2019762} prior to an exact distance computation.

A tighter (but more expensive) lower bound produces fewer candidates, while in some applications (such as DBSCAN clustering),
where the exact distance is not necessary to have, an upper bound can identify true positives efficiently.

\section{Experimental Evaluation}
\label{sec:evaluation}
In this section, we compare \textit{EmbAssi} to state-of-the-art approaches regarding efficiency and approximation quality in
range and $k$-nearest neighbor queries.
We investigate the speed-up of existing filter-verification pipelines when \textit{EmbAssi} is used in a pre-filtering step. 
Specifically, we address the following research questions:

\begin{itemize}
	\item[\textbf{Q1}] How tight are our lower bounds compared to the state-of-the-art? How do our bounds perform when taking the trade-off between bound quality and runtime into account?
	\item[\textbf{Q2}] Can \textit{EmbAssi} compete with state-of-the-art methods in terms of runtime and selectivity? Is $\CLB$ a suitable lower bound to provide initial candidates for range queries?
	\item[\textbf{Q3}] Can \textit{EmbAssi} perform similarity search on datasets with a million graphs or more?
	\item[\textbf{Q4}] Can $k$-nearest neighbor queries be answered efficiently?
\end{itemize}

\begin{figure*}[tb]
	\centering
	\includegraphics[width=\linewidth]{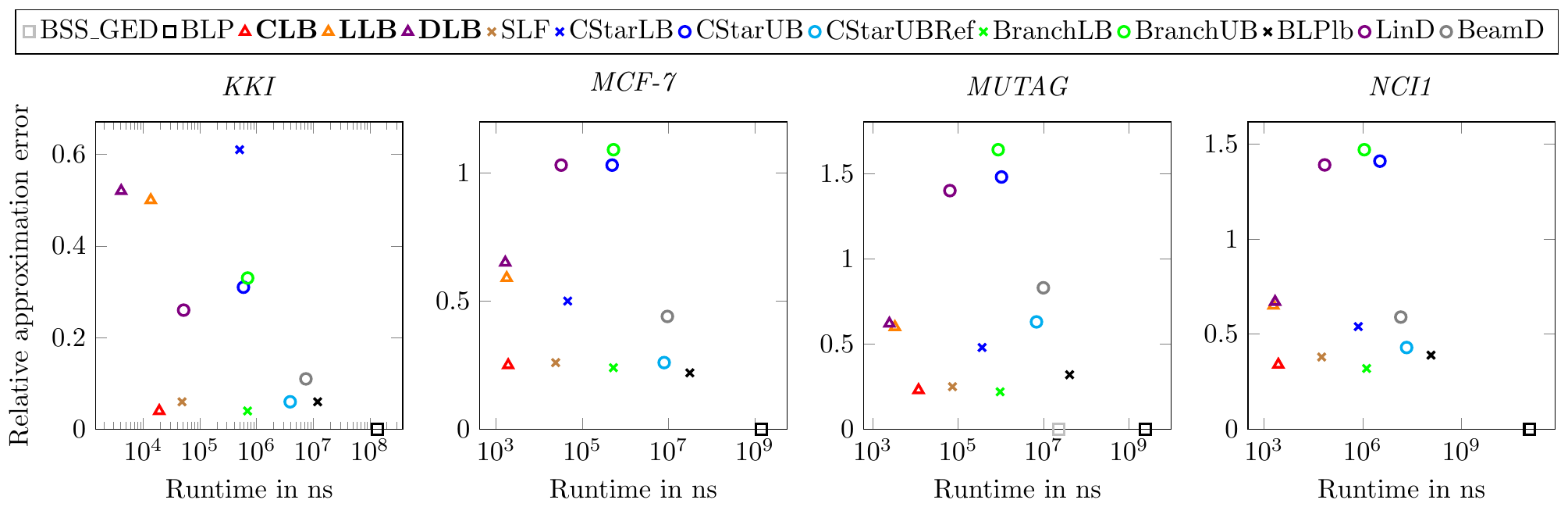}
	\caption{Comparison of several different approximations regarding their relative approximation error and runtime.
		\newline {\scriptsize$\square$}: exact approach, $\circ$: upper bound, {$\times$}: existing lower bound, {\scriptsize$\triangle$}: newly proposed lower bound from tree metric. }
	\label{fig:LZapprox}
	\vspace{-1\abovecaptionskip} %
\end{figure*}

\subsection{Setup}
\label{subsec:setup}
This section gives an overview of the datasets, the methods, and their configuration used in the experimental comparison.

\textit{Methods and Distance Functions.}
We compare \textit{EmbAssi} to \textit{GSim}~\cite{32_GSimJoin} and \textit{MLIndex}~\cite{5_simMultiIndex}, which are representative methods for similarity search based on overlapping substructures and graph partitioning. \textit{MLIndex} is considered as state-of-the-art \cite{QinBS20}, although we observed that \textit{GSim} often performs much better.
We also compare to \textit{CStar}~\cite{2_ComparingStars} and \textit{Branch}~\cite{29_ged_heuristics}, which provide both upper and lower bounds on the graph edit distance, but are not accelerated with indices.
Furthermore, we compare to the exact graph edit distance \textit{BLP}~\cite{34_blp}, \textit{BSS\_GED}~\cite{CHEN2019762}, and the approximations \textit{LinD}~\cite{1_gedLinear}, \textit{BLPlb}~\cite{29_ged_heuristics} and \textit{BeamS}~\cite{35_beamS} regarding the approximation of the \textit{GED}.
The costs of all edit operations were set to one because some of the comparison methods only support uniform costs.
For \textit{BeamD}, we used a maximum list size of 100.
For \textit{LinD}, the tree was generated using the Weisfeiler-Lehman algorithm with one refinement iteration.
For \textit{GSim}, we used all provided filters and~$q=3$.
For \textit{MLIndex}, the default settings of the authors' implementation were used.

\begin{table}[tb]\centering
	\caption{Distance functions compared in the experiments}
	\label{table:distfunc}
	\setlength{\tabcolsep}{3pt}
	\begin{tabular}{llcc}
		\toprule
		\textbf{Name} & \textbf{Description} &  \textbf{Bound}  & \textbf{Source}\\
		\midrule
		\textit{BLP} & Exact graph edit distance &  exact & \cite{34_blp}\\
		\textit{BSS\_GED} &  Efficient ged computation/verification &exact &\cite{CHEN2019762}\\
		\textit{BeamD} & Approximation using BeamSearch &  upper & \cite{35_beamS}\\
		\textit{LinD} & Optimal assignments with WL  &  upper & \cite{1_gedLinear}\\
		\textit{BranchUB} &Cost of edit path gained from BranchLB &upper& \cite{29_ged_heuristics} \\
		\textit{CStarUB} & Optimal assignments with stars & upper & \cite{2_ComparingStars} \\
		\textit{CStarUBRef} & Refined Version of \textit{CStarUB} & upper & \cite{2_ComparingStars} \\
		\textit{CStarLB} & Mapping distance between stars & lower & \cite{2_ComparingStars} \\
		\textit{SLF} & Min. label changes& lower& \cite{21_inves2019}\\
		\textit{BranchLB} &Modification of \textit{BP} &lower& \cite{29_ged_heuristics} \\
		\textit{BLPlb} &Relaxation of \textit{BLP} &lower& \cite{29_ged_heuristics} \\
		\midrule
		\multicolumn{4}{c}{\textbf{New distance functions based on tree metrics}}\\
		\midrule
		$\LLB$ & Min. vertex label changes  & lower & new \\
		$\DLB$ & Min. edge insertion/deletions  & lower & new \\
		$\CLB$ & $\LLB$ and $\DLB$ combined  & lower & new \\
		\bottomrule
	\end{tabular}
	\vspace{-1\abovecaptionskip} %
\end{table}

The bounds computed by \textit{CStar} can also be used separately and will be referred to as \textit{CStarLB}, \textit{CStarUB}, and \textit{CStarUBRef}, which is obtained by improving an edit path using local search.
\textit{SLF} and \textit{BranchLB} are the lower bounds discussed in Section~\ref{sec:clb} and were implemented following the description in \cite{21_inves2019} and \cite{29_ged_heuristics}, respectively. \textit{BranchLB} and the upper bound gained from the edit path induced by it (\textit{BranchUB}) are referred to as \textit{Branch} when applied together.
Table~\ref{table:distfunc} gives an overview of the distance functions compared in the experiments including those known from literature as well as those proposed here for use with \textit{EmbAssi}. The graphs and their respective vectors are indexed
using the cover tree~\cite{BeygelzimerKL06} implementation of the ELKI framework~\cite{DBLP:journals/corr/abs-1902-03616}.

\textit{Datasets.}
\begin{table}[tb]\centering
	\caption{Datasets with discrete vertex labels and their statistics~\cite{Datasets}. 
		\textit{ChEMBL} (\textit{chembl\_27}, \cite{Chembl}) contains small molecules, \textit{Protein Com} contains protein complex graphs~\cite{Stoecker2019}.}
	\label{tab:datasets}
	\begin{tabular}{lrrrrr}
		\toprule
		\textbf{Name} & \boldmath\textbf{$\vert$Graphs$\vert$} & \boldmath\textbf{avg $\vert V\vert$} & \boldmath\textbf{avg $\vert E \vert$} & \boldmath\textbf{avg $\delta(v)$} & \boldmath\textbf{$\vert L\vert$}\\
		\midrule
		\textit{KKI}	&	$83$ & $26.96$&	$48.42$ & $ 3.59$ \scriptsize{$\pm 2.58$} & $190$\\
		\textit{MCF-7}	&	$27770$&	$26.08$&	$28.29$ &  $2.16$ \scriptsize{$\pm 0.76$} & $23$\\
		\textit{MUTAG}&	$188$&	$17.93$&	$19.79$ & $2.20$ \scriptsize{$\pm 0.74$} & $7$\\	
		\textit{NCI1} &	$4110$& $29.04$&	$31.61$ & $2.16$ \scriptsize{$\pm 0.78$} &$21$\\
		\textit{PTC\_FM}&	$349$ &	$14.11$&	$14.48$ & $2.05$ \scriptsize{$\pm 0.81$}&$18$\\
		\midrule
		\textit{Protein Com} & $1455324$ & $9.98$ & $8.98$ & $1.80$ \scriptsize{$\pm 3.03$} &$717$\\
		\textit{ChEMBL} & $1941411$& $30.18$&$32.71$ & $2.15$ \scriptsize{$\pm 0.74$}& $31$\\
		\bottomrule
	\end{tabular}
	\vspace{-1\abovecaptionskip} %
\end{table}
We tested all methods on a wide range of real-world datasets with different characteristics, see Table~\ref{tab:datasets}.
The datasets have discrete vertex labels. Edge labels and attributes, if present, were removed prior to the experiments since not all methods support them. Also, since \textit{MLIndex} and \textit{GSim} do not work for disconnected graphs, only their largest connected components were used.

\begin{figure*}[tb]
	\centering
	\includegraphics[width=\textwidth]{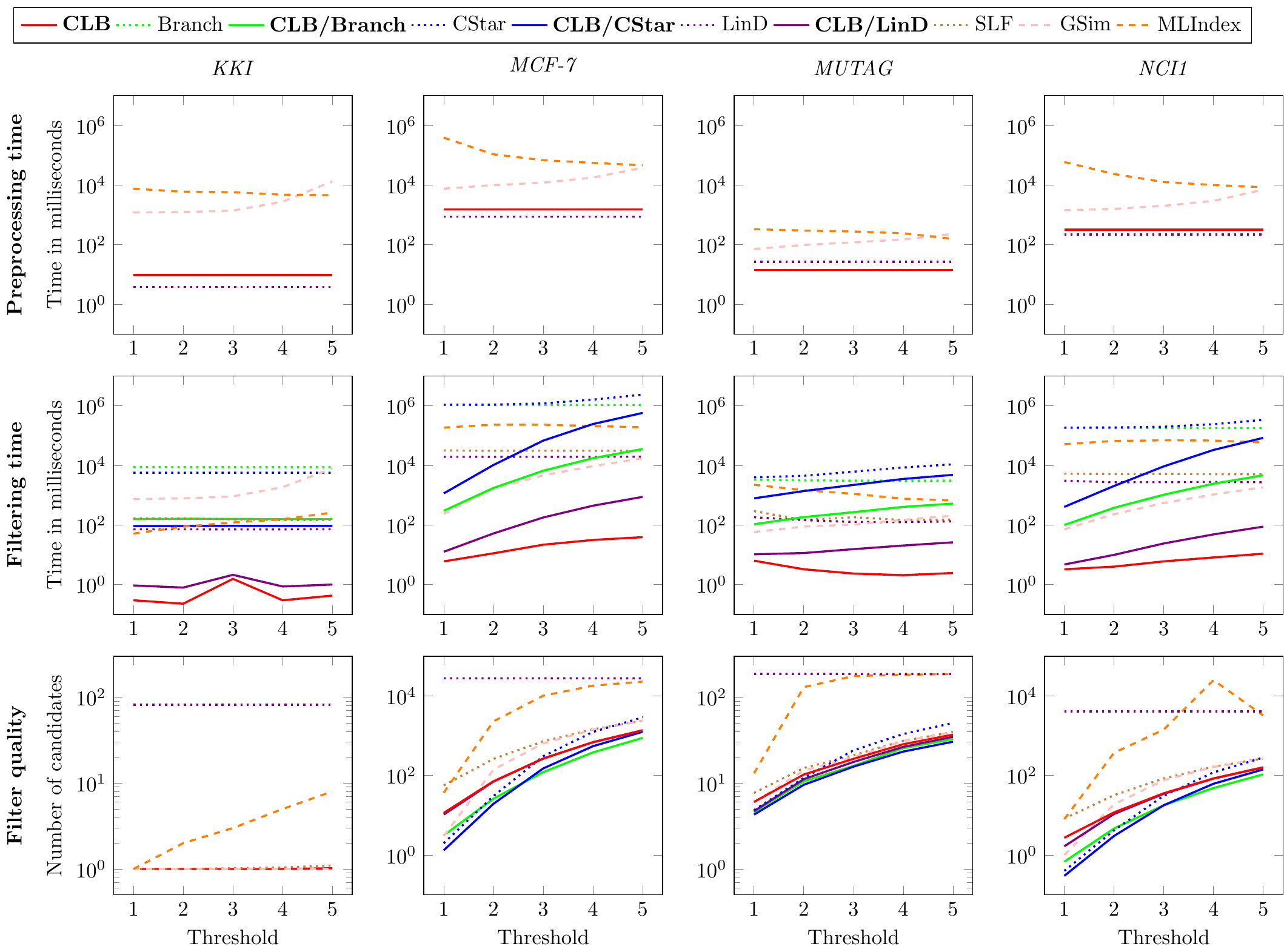}
	\caption{Runtime and selectivity comparison of different filters. Preprocessing time, filtering time for 50 range queries (excluding verification), and the average number of candidates, that need to be verified, is shown. For the methods, that can be enhanced using \textit{EmbAssi}, the solid line shows the advantage of pre-filtering with $\CLB$, while the dotted line is the original approach.}
	\label{fig:comparisonlineplot}
	\vspace{-1\abovecaptionskip} %
\end{figure*}

\begin{figure}[tb]
	\centering
	\includegraphics[width=0.75\textwidth]{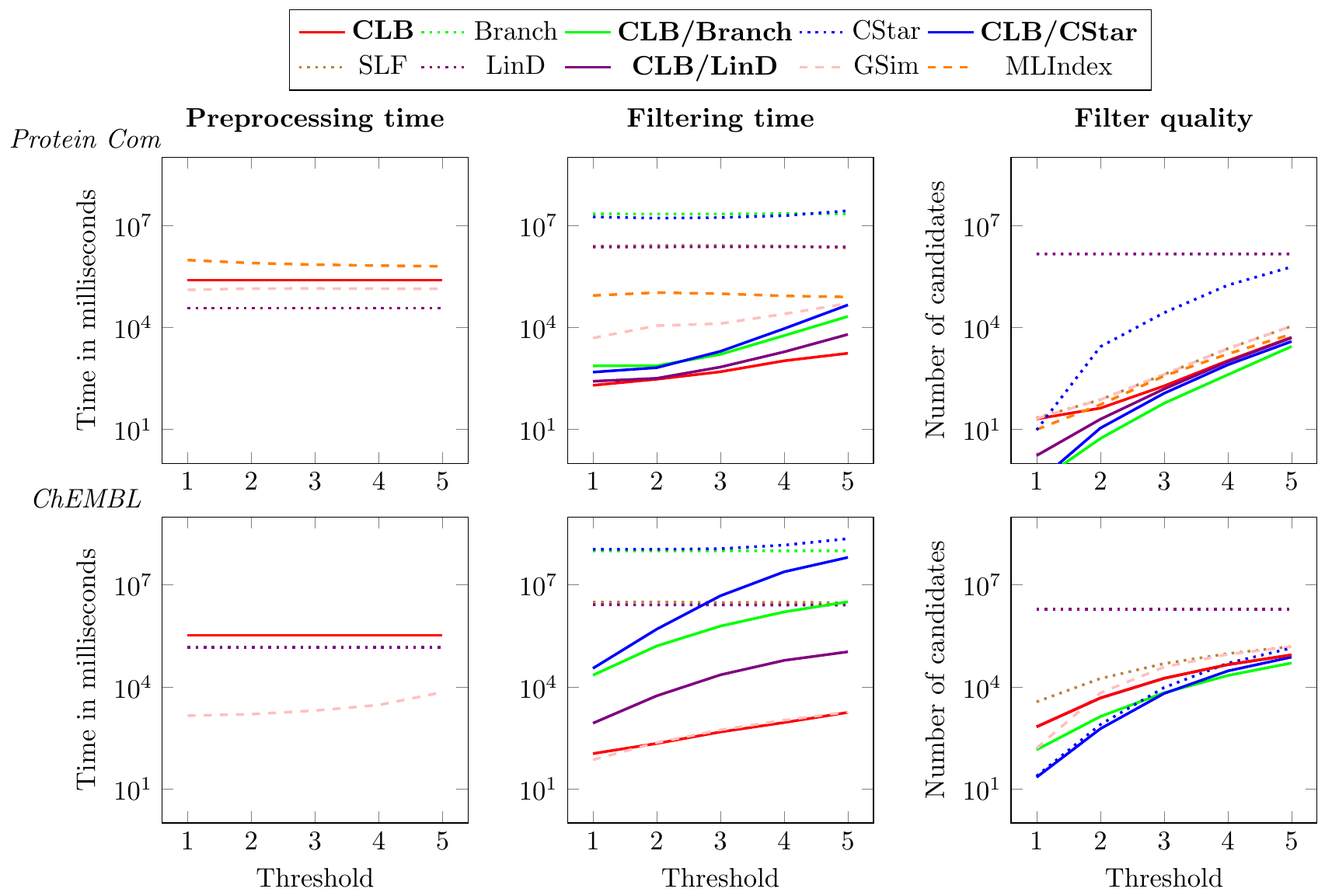}
	\caption{Comparison of several different filters regarding their selectivity and runtime on datasets \textit{Protein Com} and \textit{ChEMBL}. The solid lines show the advantage of using \textit{EmbAssi} with $\CLB$ as a pre-filter, while the dotted lines show the original approaches.
	}
	\label{fig:filterBigDS}
	\vspace{-1\abovecaptionskip} %
\end{figure}

\subsection{Results}
\label{subsec:algorithmsInComparison}

In the following, we report on our experimental results and discuss the different research questions.

\textbf{Q1: Bound Quality and Runtime.}
Accuracy is crucial to obtain effective filters for similarity search.
We investigate how tight the proposed lower bounds on the graph edit distance are.
Fig.~\ref{fig:LZapprox} shows the average relative approximation error $\frac{\abs{\text{GED} - d_{\text{approx}}}}{\text{GED}}$ of the different bounds in comparison to their runtime. The newly proposed bounds, as well as \textit{SLF}, are very fast, with varying degrees of accuracy. Although $\CLB$ is much faster than \textit{BranchLB}, its accuracy is in many cases on par or only slightly worse.
Note that a timeout of 120 seconds per graph pair was used for the computation of the exact graph edit distance for this experiment. For this reason, values for \textit{BSS\_GED} are not present for the datasets with larger graphs.

\textbf{Q2: Evaluation of Runtime and Selectivity.}
The runtime of the algorithms consists of three parts:
(1)~preprocessing and indexing, (2)~filtering, and (3)~verification.
Preprocessing and indexing is performed only once, and this cost amortizes over many queries, while the time required to determine the candidate set and its size are crucial.
The verification step requires to compute the exact graph edit distance and is usually most expensive and essentially depends on the number of candidates.

In the following, we investigate how well \textit{EmbAssi} performs on range queries, how much of a speed-up can be achieved for existing pipelines when filtering with \textit{EmbAssi} first, and compare to state-of-the-art approaches.
We omit bounds that were shown in the previous experiments to have a poor accuracy or a very high runtime.
Fig.~\ref{fig:comparisonlineplot} shows the runtime for preprocessing, filtering and the average number of candidates per query  for range queries with thresholds 1 to~5.
The solid lines show the results, when using EmbAssi with $\CLB$ as a first filter, while the dotted line represent the original approaches. The solid red line shows the results using only \textit{EmbAssi} with $\CLB$ and no further filters.
\textit{GSim} and \textit{MLIndex} are shown with dashed lines, since they are stand-alone approaches.
These two methods skip database graphs that are smaller than the given threshold. To obtain a valid candidate set, these graphs were added back after filtering.
For \textit{GSim} and \textit{MLIndex} the preprocessing time is rather high and highly dependent on the maximum threshold for range search, which must be chosen in advance.

It becomes evident that \textit{EmbAssi} significantly accelerates all methods across the various datasets. The preprocessing and filtering time of \mbox{\textit{EmbAssi}} is very low: While filtering only takes a few milliseconds, preprocessing ranges from 0.01 to 2 seconds over the various datasets.
$CStar$ and $Branch$ have the best selectivity, but they also employ both upper and lower bounds and need more time for filtering. The usage of \textit{EmbAssi} heavily accelerates both methods, while even increasing the selectivity of $CStar$ (as seen in Fig.~\ref{fig:LZapprox}, $CStarLB$ seems to be looser than $\CLB$ in general). Note, that \textit{LinD} is an upper bound, so the candidate set consist of all graphs, that could not be reported as a result. In combination with \textit{EmbAssi} it is only slightly worse than the other approaches regarding filter selectivity, while being very fast. 

Considering the properties of the datasets and the performance, we observe that a larger set of vertex labels and a high variance among the vertex degrees seem to lead to a better filter quality. The larger the graphs, the greater the improvement in runtime during the filtering step.

Since competing approaches do not use the fast verification algorithm \textit{BSS\_GED}~\cite{CHEN2019762} a comparison of verification time would not be fair. On the various datasets the time for verification (of 50 queries with threshold 5) using the candidates of $\CLB$ ranged from around $35$ms (KKI) to a maximum of $5$s (MCF-7).

Combining these results, we can conclude that \textit{EmbAssi} is well suited as pre-filtering for more effective computational demanding bounds. \textit{EmbAssi} substantially reduces the filtering time and promises scalability even to very large datasets. We investigate this below.

\textbf{Q3: Similarity Search on Very Large Datasets.}
We investigate how well \textit{EmbAssi} performs on very large graph databases using the datasets \textit{Protein Com} and \textit{ChEMBL}. 
Fig.~\ref{fig:filterBigDS} shows the average number of candidates per query as reported by the different methods, as well as the time needed for preprocessing and filtering.
\textit{MLIndex} did not finish on \textit{ChEMBL} within a time limit of 24 hours (for threshold~1).
For dataset \textit{Protein Com} our new approach is not only much faster, but also provides a better filter quality than state-of-the-art methods.
It can clearly be seen, that \textit{EmbAssi} with $\CLB$ provides a substantial boost in runtime, while also improving the filter quality. %

\textbf{Q4: \boldmath$k$-Nearest-Neighbor Search.}
An advantage of \textit{EmbAssi} is that it can also answer $k$-nn queries efficiently due to the use of the multi-step $k$-nearest neighbor search algorithm as described in Section~\ref{subsec:database}.
Table~\ref{table:knn} compares the average number of candidates generated using \textit{EmbAssi} (with $\CLB$) and \textit{BranchLB}, as well as the average time needed for answering a $k$-nn query.
In both methods, candidate sets were verified using the faster exact graph edit distance computation \textit{BSS\_GED}.
The last column shows the average number of nearest neighbors reported, which may be larger than $k$ because of ties.

\begin{table}[tb]\centering
	\caption{Runtime and number of candidates in $k$-nearest-neighbor search using \textit{EmbAssi} and \textit{BranchLB}.} 
	\label{table:knn}
	\begin{tabular}{lcrrrrr}
		\toprule
		&\multirow{2}{*}{\boldmath$k$}  & \multicolumn{2}{c}{\textbf{EmbAssi}} &\multicolumn{2}{c}{\textbf{BranchLB}} & \multirow{2}{*}{\boldmath\textbf{$\vert$NN$\vert$}}\\ \cmidrule{3-6}
		&& {$\vert$Cand$\vert$} & {Time (sec)} &{$\vert$Cand$\vert$}& {Time (sec)} &  \\
		\midrule
		\multirow{5}{*}{\rotatebox{90}{\textbf{PTC\_FM}}}
		&1 & 14.40 & 0.24 &  9.00 & 1.42 & 1.20\\
		&2 & 24.40 & 0.42 & 19.60 & 1.34 & 3.40\\
		&3 & 28.40 & 0.43 & 22.20 & 1.16 & 4.40\\
		&4 & 31.80 & 1.06 & 25.20 & 1.40 & 5.60\\
		&5 & 39.00 & 1.19 & 31.20 & 1.24 & 6.60\\
		\midrule
		\multirow{5}{*}{\rotatebox{90}{\textbf{MUTAG}}}
		&1 &  6.80 & 0.15 &  6.80 & 0.87 & 1.60\\
		&2 &  9.80 & 0.22 &  9.80 & 0.99 & 3.80\\
		&3 & 14.00 & 0.25 & 14.00 & 1.31 & 5.80\\
		&4 & 14.00 & 0.49 & 14.00 & 1.03 & 5.80\\
		&5 & 15.80 & 0.55 & 15.80 & 1.08 & 7.40\\
		\midrule
		\multirow{5}{*}{\rotatebox{90}{\textbf{MCF-7}}}
		&1 &  659.67 &  48.04 &  434.33 & 191.95 & 1.33\\
		&2 & 1114.33 & 152.43 &  737.00 & 244.42 & 2.33\\
		&3 & 1610.00 & 214.64 & 1045.67 & 410.11 & 4.00\\
		&4 & 1968.33 & 391.37 & 1380.33 & 577.75 & 10.00\\
		&5 & 2401.00 & 642.29 & 1696.33 & 678.06 & 10.67\\
		\bottomrule
	\end{tabular} 
	\vspace{-1\abovecaptionskip} %
\end{table}

It can be seen, that \textit{EmbAssi} provides a runtime advantage in $k$-nearest neighbor search, and the number of candidates generated is not much higher than when using \textit{BranchLB}. For larger datasets, we expect the advantage of \textit{EmbAssi} to be more significant. Further optimization of the approach is possible. For example, it might be beneficial to combine both methods and use \textit{EmbAssi} in combination with tighter lower bounds such as \textit{BranchLB} to reduce the number of exact graph edit distance computations.

\section{Conclusions}
\label{sec:conclusions}
We have proposed new lower bounds on the graph edit distance, which are efficiently computed, readily combined with indices, and fairly selective in filtering. This makes them ideally suitable as a pre-filtering step in existing filter-verification pipelines that do not scale to large databases.
Our approach supports efficient $k$-nearest neighbor search using the optimal multi-step $k$-nearest neighbor search algorithm unlike many comparable methods.
Other methods have to first perform a range query with a sufficient range and find the $k$-nearest neighbors among those candidates.

An interesting direction of future work is the combination and development of indices for computational demanding lower bounds such as those obtained from general assignment problems or linear programming relaxations.
Efficient methods for similarity search regarding the Wasserstein distance have only recently been investigated~\cite{30_scalableNNS}. 
Moreover, approximate filter techniques for the graph edit distance based on embeddings learned by graph neural networks were only recently proposed \cite{QinBS20}. With the increasing amount of structured data, scalability is a key issue in graph similarity search.

\setcitestyle{numbers}
\bibliography{lit.bib}

\end{document}